\documentclass[epj,nopacs]{svjour}
\usepackage{graphicx}
\newcommand{\be}{\begin{equation}}
\newcommand{\ee}{\end{equation}}
\newcommand{\bea}{\begin{eqnarray}}
\newcommand{\eea}{\end{eqnarray}}

\begin{document}
\title{Radiative Breaking Scenario for the GUT gauge symmetry}
\author{Takeshi Fukuyama\inst{1} \and Tatsuru Kikuchi\inst{1}
}                     
%
%
\institute{Department of Physics, Ritsumeikan University, 
Kusatsu, Shiga, 525-8577 Japan \\
\email{fukuyama@se.ritsumei.ac.jp}\\
\email{rp009979@se.ritsumei.ac.jp}}
\date{\today}
%
\abstract{
The origin of the GUT scale from the top down perspective 
is explored. The GUT gauge symmetry is broken by the renormalization 
group effects, which is an extension of the radiative electroweak 
symmetry breaking scenario to the GUT models. 
That is, in the same way as the origin of the electroweak scale, 
the GUT scale is generated from the Planck scale through the radiative 
corrections to the soft SUSY breaking mass parameters. 
This mechanism is applied to a perturbative SO(10) GUT model, recently 
proposed by us. In the SO(10) model, the relation between the GUT scale 
and the Planck scale can naturally be realized by using order one coupling 
constants. 
\PACS{
      {12.10.-g}{}   \and
      {12.10.Dm}{}   \and
      {12.10.Kt}{}
     } 
} 
\maketitle
\section{Introduction}
A particularly attractive idea for the physics beyond the standard model (SM)
is the possible appearance of grand unified theory (GUT) \cite{gut}. 
The idea of GUTs bears several profound features. 
Perhaps the most obvious one is that GUTs have the potential to unify 
the diverse set of particle representations and parameters found in 
the SM into a single, comprehensive, and hopefully predictive framework. 
For example, through the GUT symmetry one might hope to explain 
the quantum numbers of the fermion spectrum, or even the origins of 
fermion mass.  Moreover, by unifying all $U(1)$ generators within a 
non-Abelian theory, GUT would also provide an explanation for the 
quantization of electric charge. By combining GUT with supersymmetry (SUSY), 
we hope to unify the attractive features of GUT simultaneously with those of 
SUSY into a single theory, SUSY GUT \cite{susygut}. The apparent gauge 
couplings unification of the minimal supersymmetric standard model (MSSM) 
is strong circumstantial evidence in favor of the emergence of a SUSY GUT 
near $M_{\rm GUT} \simeq 2 \times 10^{16}$ [GeV] \cite{susygut2-1} 
\cite{susygut2-2}. 

While there are many appealing features in SUSY GUT, 
from more fundamental theory point of view, 
it looks like a sort of a problem that the discrepancy between 
the fundamental scale, say, the (reduced) Planck scale, 
$M_{\rm Pl} \simeq 2.4 \times 10^{18}$ [GeV] and the GUT scale, 
$M_{\rm GUT} \simeq 2 \times 10^{16}$ [GeV]. 
There has already been many approaches to this problem. 
Recent development of the extra dimensional physics may provide one of 
the solutions \cite{Dienes:1996du}. 
That says that the discrepancy is a consequence of a distortion of the 
renormalization group (RG) runnings by the change of space dimensions, 
and the true GUT scale would be raised up to the Planck scale. Though 
it is interesting, here we seek for the other approaches. That is, 
the ``dynamical'' generation of the GUT scale. Namely, we assume 
the theory has no any dimensionful parameters at the beginning except 
for the Planck scale. The GUT scale, what we call it from the low energy 
perspective, is generated from the radiative corrections. That lifts 
the flatness of the original potential. Then the GUT scale emerges from 
the dynamics. This has already been used to break the electroweak gauge 
symmetry \cite{Inoue:1982pi} \cite{Ibanez:1982fr} \cite{Ibanez:1983wi} 
\cite{Alvarez-Gaume:1983gj} \cite{Ellis:1982wr} \cite{Ellis:1983bp} 
that may be regarded as the first evidence of some supersymmetric 
extensions of the standard model. 

In this paper, we follow the same idea but extend the gauge group of 
the electroweak theory $SU(2)_L \times U(1)_Y$ to a simple gauge group 
for the GUT, {\it e.g.}~$SO(10)$, 
and consider to break it to the Standard Model one, 
$SU(3)_c \times SU(2)_L \times U(1)_Y$ by radiative corrections. 
Indeed, the soft SUSY breaking mass parameters for the GUT Higgs multiplet 
can be driven to the negative values at the GUT scale through their RG 
runnings \cite{Gato:1983yz} \cite{Gato:1984ya} \cite{Yamamoto:1983yz} 
\cite{Goldberg} \cite{Bajc}. 
We explicitly construct a model with such a radiative GUT breaking 
scenario, and apply it to the $SO(10)$ model that allows perturbative 
calculations up to the Planck scale and satisfies low energy phenomena 
\cite{Chang}. Then the GUT scale is determined only by the order one 
Yukawa couplings and the Planck scale. This deep connection between 
the GUT scale and the Planck scale leads us to believe the theory of 
grand unification. 

\section{Toy model}
First we consider an $SU(5)$ GUT model discussed in \cite{Goldberg}, 
clarifying the argument. Let us denote $S$, $H$, $\overline{H}$ and $\Sigma$ 
an $SU(5)$ singlet, fundamental, anti-fundamental and the GUT breaking 
adjoint Higgs superfields, respectively. 
By giving global $U(1)$ charges for these fields as 
$S=+2$, $H= \overline{H} = +1/2$, $\Sigma=-1$, 
we have a superpotential of the form: 
\be
W(S, \Sigma)=  \lambda_\Sigma S\, {\rm Tr}(\Sigma^2)
+ \lambda_H \overline{H} \Sigma H \;.
\label{W}
\ee
Obviously, from Eq.~(\ref{W}) we get 
\bea
F_\Sigma^\dag &=& \frac{\partial{W}}{\partial{\Sigma}} 
= 2 \lambda_\Sigma S \Sigma + \lambda_H \overline{H} H = 0\;,
\nonumber\\
F_S^\dag &=& \frac{\partial{W}}{\partial{S}} 
= \lambda_\Sigma \,{\rm Tr}(\Sigma^2) = 0 \;.
\eea
That leads to one of the vacua:
$\left< \Sigma \right> =0$, $\left<S \right> = {\rm arbitrary}$, 
regarding the electroweak scale VEV's as zero: 
$\left< H \right> = \left< \overline{H} \right> =0$. 
For the true vacuum which respects the SM gauge symmetry 
$SU(3)_c \times SU(2)_L \times U(1)_Y$, we want to have a non-zero VEV 
for the adjoint Higgs field in the following direction:
\footnote{Though there is an equivalent possibility to have a VEV in 
the other direction $SU(4) \times U(1) \subset SU(5)$, 
here we just take the VEV in the desirable SM direction by hand.}
\be
\left< \Sigma \right> = \frac{1}{\sqrt{30}}{\rm diag}(2,2,2,-3,-3) \, v~,v \neq 0 \;.
\ee
Here we include the soft SUSY breaking mass terms,
\be
V_{\rm soft} = m_\Sigma^2 |\Sigma|^2 + m_S^2 |S|^2 
+ \frac{1}{2} M_\lambda \lambda_a^T C^{-1} \lambda_a \;,
\ee
where $\lambda_a$ $(a=1, \cdots, 24)$ is the $SU(5)$ gaugino. 
Taking this into account, the total scalar potential is given by 
\bea
V &=& m_\Sigma^2 |\Sigma|^2 + m_S^2 |S|^2 
+ \left|2 \lambda_\Sigma S \Sigma \right|^2
+ \left|\lambda_\Sigma \,{\rm Tr}(\Sigma^2) \right|^2 
\nonumber\\
&+& \frac{1}{2} M_\lambda \lambda_a^T C^{-1} \lambda_a \;.
\eea
Then the potential minima can be obtained as follows: 
\bea
\frac{\partial{V}}{\partial{\Sigma^\dag}} 
&=& \left(m_\Sigma^2 + 4 \lambda_\Sigma^2 S^2 \right) \Sigma
\ =\ 0 \;,\\
\frac{\partial{V}}{\partial{S^\dag}} 
&=& \left(m_S^2 + 4 \lambda_\Sigma^2 \,{\rm Tr}(\Sigma^2) \right) S
\ =\ 0 \;,
\eea
that is, one of the vacua which respects the SM gauge symmetry is found to be 
\bea
&&\left< S \right> \simeq \sqrt{\frac{-m_\Sigma^2}{4 \lambda_\Sigma^2}}\;,
\nonumber\\
&&\left< \Sigma \right> = \frac{1}{\sqrt{30}}{\rm diag}(2,2,2,-3,-3)\, v \;,~~
v \simeq \sqrt{\frac{-m_S^2}{4 \lambda_\Sigma^2}} \;.\quad
\eea
Here the negative mass squared for the singlet, $m_S^2 < 0$ should be 
satisfied at the GUT scale to realize the correct symmetry breaking. 
In the following, we really see that such a negative value can be 
achieved through the RG running from the Planck scale to the GUT scale 
with a large enough Yukawa coupling even if we start with a positive 
mass squared at the Planck scale. 

\section{RG analysis}
The RG equations for the Yukawa couplings and the soft SUSY breaking 
mass terms are given by \cite{Martin} \cite{Yamada:1994id} 
\bea
16 \pi^2 \mu\frac{d\lambda_\Sigma}{d\mu}&=& 
\left(14 \lambda_\Sigma^2 - 20 g^2 \right) \lambda_\Sigma \;,
\nonumber\\
16 \pi^2 \mu\frac{dm_S^2}{d\mu}&=& 
24 \lambda_\Sigma^2 \left(m_S^2 + 2 m_\Sigma^2 \right)\;,
\nonumber\\
16 \pi^2 \mu\frac{dm_\Sigma^2}{d\mu}&=& 
2 \lambda_\Sigma^2 \left(m_S^2 + 2 m_\Sigma^2 \right) - 40 M_a^2 \;,
\nonumber\\
16 \pi^2 \mu\frac{dM_a}{d\mu}&=& - 20 g^2 M_a \;,
\nonumber\\
16 \pi^2 \mu\frac{dg}{d\mu} &=& - 10 g^3 \;.
\eea
In the limit of the vanishing gaugino 
masses, that is, in the exact ${\cal R}$ symmetric limit, 
the RG equations for the soft SUSY breaking scalar masses 
lead to the following forms:
\bea
\mu\frac{dm_S^2}{d\mu}&=& 
\frac{3}{2 \pi^2} \lambda_\Sigma^2 \left(m_S^2 + 2 m_\Sigma^2  \right)\;,
\nonumber\\
\mu\frac{dm_\Sigma^2}{d\mu}&=& 
\frac{1}{8 \pi^2} \lambda_\Sigma^2 \left(m_S^2 + 2 m_\Sigma^2   \right)\;.
\label{RG}
\eea
Here we have assumed the coupling constant $\lambda_\Sigma$ being a constant 
number against the renormalization from $M_{\rm Pl}$ to $M_{\rm GUT}$. 
One combination leads to the following equation, 
\be
\mu\frac{d}{d\mu}\left(m_S^2 + 2 m_\Sigma^2 \right) = 
\frac{7}{4 \pi^2} \lambda_\Sigma^2 \left(m_S^2 + 2 m_\Sigma^2 \right)\;.
\label{EQ}
\ee
Assuming the universal soft mass parameter $m_{3/2}$ at the Planck scale, 
the solution is found to be: 
\be
m_S^2 + 2 m_\Sigma^2 = 3 m_{3/2}^2 \exp \left[
\frac{7}{4 \pi^2} \lambda_\Sigma^2 
\ln \left(\frac{\mu}{M_{\rm Pl}} \right) \right]\;.
\label{sol}
\ee
It gives a solution for $m_S^2$ as follows: 
\be
m_S^2 = - \frac{11}{7} m_{3/2}^2 
+ \frac{18}{7} m_{3/2}^2 \exp \left[\frac{7}{4 \pi^2} \lambda_\Sigma^2 
\ln \left(\frac{\mu}{M_{\rm Pl}} \right) \right]\;,
\label{mS}
\ee
and also the solution for $m_\Sigma^2$ is found to be 
\be
m_\Sigma^2 = \frac{11}{14} m_{3/2}^2 
+ \frac{3}{14} m_{3/2}^2 \exp \left[\frac{7}{4 \pi^2} \lambda_\Sigma^2 
\ln \left(\frac{\mu}{M_{\rm Pl}} \right) \right]\;.
\label{mA}
\ee
The solution for $m_S^2$ has two opposite sign terms as you can see in 
Eq.~(\ref{mS}), and the RG running from the Planck scale may drive it 
to the negative value to induce the GUT symmetry breaking. Here we show 
the relation explicitly. The required condition to achieve the radiative 
GUT symmetry breaking is $m_S^2 = 0$ at a scale $\mu = M_{\rm GUT}$. 
From this requirement, the GUT scale is generated from the Planck 
scale via the dimensional transmutation: 
\be
M_{\rm GUT} = M_{\rm Pl}
\exp\left[\frac{4 \pi^2}{7 \lambda_\Sigma^2}
\ln \left(\frac{11}{18} \right)
\right]\;.
\label{GUT}
\ee
One of the important things is that this expression depicting 
the GUT scale is completely independent of the SUSY breaking scale 
$m_{3/2}$, and the order one coupling constant ($\lambda_\Sigma \simeq 0.72$) 
can realize the appropriate GUT scale 
$M_{\rm GUT} \simeq 2 \times 10^{16}$ [GeV]. 
This completely reflects the situation similar to the radiative electroweak 
symmetry breaking scenario \cite{Inoue:1982pi} \cite{Ibanez:1982fr} 
\cite{Ibanez:1983wi} \cite{Alvarez-Gaume:1983gj} \cite{Ellis:1982wr} 
\cite{Ellis:1983bp}. 
This scenario postulates that the electroweak scale ($v \simeq 174$ [GeV]) 
is a consequence of the large top Yukawa coupling 
($y_t \simeq m_t/v \simeq 1.02$) which makes one of the soft SUSY breaking 
mass parameters for the Higgs doublets ($H_u$, $H_d$) bending to 
the negative value ($m_{H_u}^2 < 0$ at $\mu = M_{\rm EW}$) 
through the RG running from the Planck scale at which the soft SUSY breaking 
mass parameters are assumed to be positive 
($m_{H_u}^2 > 0$ at $\mu = M_{\rm Pl}$). 

\section{SO(10) model}
Now we proceed to extend the model into the realistic $SO(10)$ GUT model, 
in which the adjoint representation is necessary to break the $SO(10)$ 
and to provide the appropriate numbers of the would-be NG boson. 
For details of symmetry breaking patterns in $SO(10)$ models, see
\cite{Fukuyama:2004ps}.
The use of $A={\bf 45}$ representation of the Higgs field is also 
economical for the realization of the doublet-triplet splittings 
in the $SO(10)$ GUT with the help of the Dimopoulos-Wilczek mechanism 
\cite{Dimopoulos}. From Eq.~(\ref{GUT}), the coupling constant 
$\lambda_A$ is given by $\lambda_A \simeq 0.72$, that is a natural 
number to realize in going through the perturbative calculation 
\cite{Chang}. In this reference \cite{Chang}, we introduced 
a set of the Higgs as $\{\bf 10 + 10' + 45 + 16 + \overline{16}\}$ 
that is denoted by $H = {\bf 10}$, $H' = {\bf 10'}$, $A={\bf 45}$, 
$\psi ={\bf 16}$ and $\overline{\psi}={\bf \overline{16}}$. 
The Yukawa couplings with matter multiplet 
$\Psi_i = {\bf 16}_i ~(i=1, 2, 3)$ are given by
\be
W = Y_{10}^{ij} \Psi_i \Psi_j H         
  + \frac{1}{M_{\rm Pl}} \, Y_{45}^{ij} \Psi_i \Psi_j H' A
+ \frac{1}{M_{\rm Pl}} \, Y_{16}^{ij} \Psi_i \Psi_j \,
\overline{\psi} \overline{\psi} \;.
\label{minimal}
\ee
The first two terms are the Yukawa couplings of quarks, charged leptons, 
and Dirac neutrinos. The third term is that for heavy right-handed 
Majorana neutrinos which makes light Majorana neutrinos via the see-saw 
mechanism \cite{see-saw}. This is a minimal set of the Higgs which 
realizes the realistic fermion mass spectra and achieve the correct 
gauge symmetry breaking. In addition to it, we add one singlet, 
$S={\bf 1}$ and one {\bf 54} multiplet, $S'={\bf 54}$. 
Assuming global $U(1)$ charges $\Psi_i = -1~(i=1,2,3)$, 
$S=-2$, $S'=+2$, $A=-1$, $\psi=+1$, $\overline{\psi}=+1$, 
$H = +2$ and $H^\prime = +3$ 
for these fields, the relevant part of the superpotential 
for the GUT breaking sector is given by 
\be
W = \lambda_\psi S \overline{\psi} \psi
+ \lambda_A S' A^2 \;,
\label{W2}
\ee
and the corresponding soft mass terms are 
\bea
V_{\rm soft} &=&
m_H^2 |H|^2 + m_{H'}^2 |H'|^2 + m_{\overline{\psi}}^2 |\overline{\psi}|^2 
+m_\psi^2 |\psi|^2
\nonumber\\
&+& m_A^2 |A|^2 + m_S^2 |S|^2 + m_{S'}^2 |S'|^2 
\nonumber\\
&+& \frac{1}{2} M_\lambda \lambda_a^T C^{-1} \lambda_a \;.
\eea
These give a total scalar potential as follows:
\bea
V &=& 
m_H^2 |H|^2 + m_{H'}^2 |H'|^2 
+ m_{\overline{\psi}}^2 |\overline{\psi}|^2 +m_\psi^2 |\psi|^2
\nonumber\\
&+& \left|\lambda_\psi S \overline{\psi} \right|^2
+ \left|\lambda_\psi S \psi \right|^2 
+ \left|\lambda_\psi \overline{\psi} \psi \right|^2
\nonumber\\
&+& m_A^2 |A|^2 + m_S^2 |S|^2 + m_{S'}^2 |S'|^2 
\nonumber\\
&+& \left|2 \lambda_A S' A \right|^2 
+ \left|\lambda_A \left(A^2 - \frac{1}{10} {\rm Tr}(A^2) {\bf 1}
\right) \right|^2 
\nonumber\\
&+& \frac{1}{2} M_\lambda \lambda_a^T C^{-1} \lambda_a \;,
\eea
and one of the vacua which respects the SM gauge symmetry is
\bea
&&\left<H \right> = \left< H' \right> = 0\;,~
\left< \psi \right> = \left< \overline{\psi} \right> \simeq
\sqrt{\frac{-m_S^2}{\lambda_\psi^2}}\;,~
\left< S \right> \simeq \sqrt{\frac{-m_\psi^2}{\lambda_\psi^2}}\;,
\nonumber\\
&&\left< S' \right> = \frac{1}{\sqrt{60}} \left(
\begin{array}{c}
1 \ 0 \\
0 \ 1 \\
\end{array}
\right)
\otimes {\rm diag}(2,2,2,-3,-3)\,s' \;,~
s' \simeq \sqrt{\frac{-m_A^2}{4\lambda_A^2}}\;,
\nonumber\\
&&\left< A \right> = \frac{1}{\sqrt{10}} \left(
\begin{array}{cc}
0  \ 1 \\
-1 \ 0 \\
\end{array}
\right)
\otimes {\rm diag}(1,1,1,1,1)\,a \;,~
a \simeq \sqrt{\frac{-m_{S'}^2}{4\lambda_A^2}} \;.
\nonumber\\
\eea
The RG equations for the soft SUSY breaking parameters 
(in the limit of vanishing gaugino masses) are given by 
\cite{Martin} \cite{Yamada:1994id}
\footnote{In principle, these results can be read off from 
the results of \cite{Machcek} for a general gauge theory.
}
\bea
16 \pi^2 \mu\frac{dm_S^2}{d\mu} &=& 
16 \lambda_\psi^2 \left(m_S^2 + 2 m_\psi^2 \right) \;,
\nonumber\\
16 \pi^2 \mu\frac{dm_\psi^2}{d\mu} &=& 
2 \lambda_\psi^2 \left(m_S^2 + 2 m_\psi^2   \right)\;,
\nonumber\\
16 \pi \mu\frac{dm_{S'}^2}{d\mu} &=& 
45 \lambda_A^2 \left(m_{S'}^2 + 2 m_A^2 \right) \;,
\nonumber\\
16 \pi^2 \mu\frac{dm_A^2}{d\mu} &=& 
54 \lambda_A^2 \left(m_{S'}^2 + 2 m_A^2   \right)\;.
\label{RG2}
\eea
Then the solutions for $m_S^2$, $m_{\psi}^2$, $m_{S'}^2$, 
and $m_{A}^2$ are now found to be
\bea
m_S^2 &=& - \frac{7}{5} m_{3/2}^2 
+ \frac{12}{5} m_{3/2}^2 
\exp \left[\frac{5 \lambda_\psi^2 }{4 \pi^2} 
\ln \left(\frac{\mu}{M_{\rm Pl}} \right) \right]\,,
\label{sol1}
\\
m_{\psi}^2 &=& \frac{7}{10} m_{3/2}^2 
+ \frac{3}{10} m_{3/2}^2 
\exp \left[\frac{5 \lambda_\psi^2 }{4 \pi^2} 
\ln \left(\frac{\mu}{M_{\rm Pl}} \right) \right]\,,
\label{sol3}
\\
m_{S'}^2 &=& \frac{2}{17} m_{3/2}^2 
+ \frac{15}{17} m_{3/2}^2 
\exp \left[\frac{153 \lambda_A^2 }{16 \pi^2} 
\ln \left(\frac{\mu}{M_{\rm Pl}} \right) \right]\,,
\label{sol2}
\nonumber\\
\\
m_{A}^2 &=& - \frac{1}{17} m_{3/2}^2 
+ \frac{18}{17} m_{3/2}^2 
\exp \left[\frac{153 \lambda_A^2 }{16 \pi^2} 
\ln \left(\frac{\mu}{M_{\rm Pl}} \right) \right]\,.
\label{sol4}
\nonumber\\
\eea
In the $SO(10)$ case, we have two opposite sign terms for 
$m_S^2$ and $m_A^2$, and the RG runnings from the Planck scale 
drive both of them to the negative values to induce the GUT symmetry breaking 
but also to break the rank of $SO(10)$, which is necessary to realize 
the Standard Model gauge group. 
Here we remember the decompositions of each representations 
under the subgroups of $SO(10)$:
\bea
&&{\bf 54} = ({\bf 1,1,1}) + ({\bf 1,3,3}) + ({\bf 20',1,1}) + ({\bf 6,2,2})
\nonumber\\
&&\mbox{\sf under~${\sf SU(4)}_c \times {\sf SU(2)}_L \times {\sf SU(2)}_R$} 
\;, \\
&&{\bf 16} = ({\bf 4,2, 1}) + ({\bf \overline{4},1,2})
\nonumber\\
&&\mbox{\sf under~${\sf SU(4)}_c \times {\sf SU(2)}_L \times {\sf SU(2)}_R$}
\nonumber\\
&&= \left[({\bf 3,2},1/6) + ({\bf 1,2},-1/2)\right]
\nonumber\\
&&+ \left[({\bf \overline{3},1},1/3) + ({\bf \overline{3},1},-2/3) 
+ ({\bf 1,1},1) +({\bf 1,1},0) \right]
\nonumber\\
&&\mbox{\sf under~${\sf SU(3)}_c \times {\sf SU(2)}_L 
\times {\sf U(1)}_Y$} \;.
\eea
Hence, if $({\bf 1,1,1}) \in {\bf 54}$ develops a VEV, 
the gauge symmetry breaks down to $SU(4)_c \times SU(2)_L \times SU(2)_R$, 
and further if $({\bf 1,1},0) \in {\bf 16}$ develops a VEV, 
the gauge symmetry breaks down to the Standard Model one.

Eq.~(\ref{RG2}) shows that the RG effects for the coupled system of 
$\{S,~\psi,~\overline{\psi} \}$ are stronger than that of $\{S',~A \}$. 
So, combining with the above discussion this fact implies 
that $SO(10)$ first breaks to 
$SU(4)_c \times SU(2)_L \times SU(2)_R$ via $\left<S' \right>$ 
and soon breaks to $SU(3)_c \times SU(2)_L \times U(1)_Y$ via 
$\left<\psi \right>$, 
leading to the scenario that $SO(10)$ breaks to the MSSM at the GUT scale. 
It should be remarked that this breaking pattern does not so much depend 
on the choice of the coupling constants for most parameters region 
because of their large charge difference between 
$\{\bf 1 \oplus 16 \oplus \overline{16}\}$ and $\{\bf 54 \oplus 45\}$.
Note that the VEV of $\left<\psi \right>$ breaks $B-L$ and gives 
masses to the heavy right-handed Majorana neutrinos at the same time. 
In summary, the GUT scale is completely determined only by the order one 
coupling constants $\lambda_A \sim \lambda_\psi \sim 1$. 

Finally, we determine the concrete values for these coupling constants. 
To achieve a simple, one step unification picture, we impose the condition 
that the rank breaking occurs at the same time for the GUT breaking. 
Then the required conditions to achieve the radiative GUT symmetry breaking 
become $m_A^2 = 0$ and $m_S^2 = 0$ at a scale $\mu = M_{\rm GUT}$. 
From this requirement, the GUT scale is generated from the Planck scale via 
the dimensional transmutation as in the case of $SU(5)$: 
\bea
M_{\rm GUT} &=& M_{\rm Pl}
\exp\left[\frac{16 \pi^2}{153 \lambda_A^2}
\ln \left(\frac{1}{18} \right)
\right]
\nonumber\\
&=& M_{\rm Pl}
\exp\left[\frac{4 \pi^2}{5 \lambda_\psi^2}
\ln \left(\frac{7}{12} \right)
\right] \;.
\label{GUT}
\eea
From these equations, we can determine the values of order one coupling 
constants which are necessary to realize the GUT scale of order
$M_{\rm GUT} \simeq 2 \times 10^{16}$ [GeV] as 
$\lambda_A \simeq 0.79$ and $\lambda_\psi \simeq 0.94$.

\section{Conclusion}
In this letter, we have explored the origin of the GUT scale. 
Given the universal soft mass at Planck scale, the GUT scale 
is determined in terms of the order one coupling constant. That is, 
the positive soft mass runs from the Planck scale to the GUT scale 
according to the RG equations and crosses zero at the GUT scale, which 
is exactly the same idea as the radiative electroweak symmetry breaking 
scenario in the MSSM with a suitable boundary condition at the Planck scale. 
This mechanism has been applied to a $SO(10)$ model recently proposed by us 
\cite{Chang} which is compatible with low energy phenomena. 
In this $SO(10)$ model, the GUT scale can be generated from the Planck scale 
by using order one coupling constants and the symmetry breaking pattern 
of $SO(10)$ is also specified. Thus the GUT scale is determined both by 
a top down (from the Planck scale to the GUT scale) scenario as well as 
a bottom up (from the electroweak scale to the GUT scale) scenario, 
and they coincide to each other. 
\begin{acknowledgement}
The work of T.F. is supported in part by the Grant-in-Aid for Scientific 
Research from the Ministry of Education, Science and Culture of Japan 
(\#16540269). The work of T.K. was supported by the Research Fellowship 
of the Japan Society for the Promotion of Science (\#7336).
\end{acknowledgement}


\begin{thebibliography}{99}
\bibitem{gut}
J.~C.~Pati and A.~Salam,
Phys.\ Rev.\ D {\bf 10}, 275 (1974);
H.~Georgi and S.~L.~Glashow,
Phys.\ Rev.\ Lett.\  {\bf 32}, 438 (1974).

\bibitem{susygut}
N.~Sakai,
Z.\ Phys.\ C {\bf 11}, 153 (1981);
S.~Dimopoulos and H.~Georgi,
Nucl.\ Phys.\ B {\bf 193}, 150 (1981).

\bibitem{susygut2-1}
C.~Giunti, C.~W.~Kim and U.~W.~Lee,
Mod.\ Phys.\ Lett.\ A {\bf 6} (1991) 1745;
P.~Langacker and M.~x.~Luo,
Phys.\ Rev.\ D {\bf 44}, 817 (1991);
U.~Amaldi, W.~de Boer and H.~Furstenau,
Phys.\ Lett.\ B {\bf 260}, 447 (1991).

\bibitem{susygut2-2}
As early works before LEP experiments, see: 
S.~Dimopoulos, S.~Raby and F.~Wilczek,
Phys.\ Rev.\ D {\bf 24}, 1681 (1981);
L.~E.~Iba\~n\'ez and G.~G.~Ross,
Phys.\ Lett.\ B {\bf 105}, 439 (1981);
M.~B.~Einhorn and D.~R.~T.~Jones,
Nucl.\ Phys.\ B {\bf 196}, 475 (1982);
W.~J.~Marciano and G.~Senjanovi\'c,
Phys.\ Rev.\ D {\bf 25}, 3092 (1982).

\bibitem{Dienes:1996du}
For a review, see, e.g. K.~R.~Dienes,
Phys.\ Rept.\  {\bf 287}, 447 (1997)
[arXiv:hep-th/9602045].

\bibitem{Inoue:1982pi}
K.~Inoue, A.~Kakuto, H.~Komatsu and S.~Takeshita,
Prog.\ Theor.\ Phys.\  {\bf 68}, 927 (1982)
[Erratum-ibid.\  {\bf 70}, 330 (1983)].

\bibitem{Ibanez:1982fr}
L.~E.~Iba\~n\'ez and G.~G.~Ross,
Phys.\ Lett.\ B {\bf 110}, 215 (1982).

\bibitem{Ibanez:1983wi}
L.~E.~Iba\~n\'ez and C.~Lopez,
Phys.\ Lett.\ B {\bf 126}, 54 (1983).

\bibitem{Alvarez-Gaume:1983gj}
L.~Alvarez-Gaum\'e, J.~Polchinski and M.~B.~Wise,
Nucl.\ Phys.\ B {\bf 221}, 495 (1983).

\bibitem{Ellis:1982wr}
J.~R.~Ellis, D.~V.~Nanopoulos and K.~Tamvakis,
Phys.\ Lett.\ B {\bf 121}, 123 (1983).

\bibitem{Ellis:1983bp}
J.~R.~Ellis, J.~S.~Hagelin, D.~V.~Nanopoulos and K.~Tamvakis,
Phys.\ Lett.\ B {\bf 125}, 275 (1983).

\bibitem{Gato:1983yz}
B.~Gato, J.~Leon and M.~Quiros,
Phys.\ Lett.\ B {\bf 136}, 361 (1984).

\bibitem{Gato:1984ya}
B.~Gato, J.~Leon, J.~Perez-Mercader and M.~Quiros,
Nucl.\ Phys.\ B {\bf 253}, 285 (1985).

\bibitem{Yamamoto:1983yz}
K.~Yamamoto,
Phys.\ Lett.\ B {\bf 135}, 63 (1984).

\bibitem{Goldberg}
H. Goldberg, 
Phys.\ Lett.\ B {\bf 400}, 301 (1997) 
[arXiv:hep-ph/9701373]. 
The extension to the flipped SU(5) model was given by
A. Dedes, C.~Panagiotakopoulos and K.~Tamvakis,
Phys.\ Rev.\ D {\bf 57}, 5493 (1998).

\bibitem{Bajc}
B.~Bajc, I.~Gogoladze, R.~Guevara and G.~Senjanovi\'c, 
Phys.\ Lett.\ B {\bf 525}, 189 (2002)
[arXiv:hep-ph/0108196].

\bibitem{Chang}
D.~Chang, T.~Fukuyama, Y.~Y.~Keum, T.~Kikuchi and N.~Okada,
Phys.\ Rev.\ D {\bf 71}, 095002 (2005)
[arXiv:hep-ph/0412011].

\bibitem{Martin}
S.~P.~Martin and M.~T.~Vaughn,
Phys.\ Rev.\ D {\bf 50}, 2282 (1994)
[arXiv:hep-ph/9311340].

\bibitem{Yamada:1994id}
Y.~Yamada,
Phys.\ Rev.\ D {\bf 50}, 3537 (1994)
[arXiv:hep-ph/9401241].

\bibitem{Machcek}
M.~E.~Machacek and M.~T.~Vaughn, Nucl.\ Phys.\ B {\bf 222}, 83 (1983);
{\it ibid.} Nucl.\ Phys.\ B {\bf 236}, 221 (1984);
{\it ibid.} Nucl.\ Phys.\ B {\bf 249}, 70 (1985).

\bibitem{Fukuyama:2004ps}
T.~Fukuyama, A.~Ilakovac, T.~Kikuchi, S.~Meljanac and N.~Okada,
Euro.\ Phys.\ J.\ C {\bf 42}, 191 (2005) [arXiv:hep-ph/0401213];
{\it ibid.} J.\ Math.\ Phys.\  {\bf 46}, 033505 (2005) 
[arXiv:hep-ph/0405300].

\bibitem{Dimopoulos}
S.~Dimopoulos and F.~Wilczek, NSF-ITP-82-07 (unpublished); 
K.S.~Babu and S.M.~Barr, Phys.\ Rev.\ D {\bf 48}, 5354 (1993). 

\bibitem{see-saw}
T.~Yanagida, in Proceedings of the workshop 
on the Unified Theory and Baryon Number in the Universe, 
edited by O.~Sawada and A.~Sugamoto (KEK, Tsukuba, 1979);
M.~Gell-Mann, P.~Ramond and R.~Slansky, 
in Supergravity, edited by D.~Freedman and P.~van~Nieuwenhuizen 
(North-Holland, Amsterdam, 1979); 
R.~N.~Mohapatra and G.~Senjanovi\'c, 
Phys.\ Rev.\ Lett. {\bf 44}, 912 (1980).
\end{thebibliography}
\end{document}